\documentclass[12pt,halfline,a4paper]{ouparticle}

\usepackage{graphicx}
\usepackage{amssymb}
\usepackage{caption}
\usepackage{url}
\usepackage{amsmath}
\usepackage{float}
\usepackage{color}
\usepackage{lineno}
\usepackage{gensymb}
\usepackage[normalem]{ulem}
\usepackage{ragged2e}

\graphicspath{{Color_Images/}}

\begin{document}


\title{Swimming through parameter subspaces of a simple anguilliform swimmer}

\author{
\name{Nicholas A. Battista}
\address{Dept. of Mathematics and Statistics, The College of New Jersey, 2000 Pennington Road, Ewing Township, NJ 08628, USA}
\email{battistn@tcnj.edu}
}


\abstract{
Computational scientists have investigated swimming performance across a multitude of different systems for decades. Most models depend on numerous model parameters and performance is sensitive to those parameters. In this paper, parameter subspaces are qualitatively identified in which there exists enhanced swimming performance for an idealized, simple swimming model that resembles a \textit{C. elegans}, an organism that exhibits an anguilliform mode of locomotion. The computational model uses the immersed boundary method to solve the fluid-interaction system.  The $1D$ swimmer propagates itself forward by dynamically changing its preferred body curvature. Observations indicate that the swimmer's performance appears more sensitive to fluid scale and stroke frequency, rather than variations in the velocity and acceleration of either its upstroke or downstroke as a whole. Pareto-like optimal fronts were also identified within the data for the cost of transport and swimming speed. While this methodology allows one to locate robust parameter subspaces for desired performance in a straight-forward manner, it comes at the cost of simulating orders of magnitude more simulations than traditional fluid-structure interaction studies. 
}


\date{\today}


\maketitle

%
%
%
%

\section{Introduction}
\label{sec:intro}

For decades numerous scientists have studied the swimming performance of animals across all lengths scales of the animal kingdom. Outside of (wet) laboratory settings, computational scientists have developed and used sophisticated modeling tools to explore a plethora of swimming systems \cite{Fish:2008,Borazjani:2009,Tytell:2010b,Bhalla:2013a,Hamlet:2015,Daghooghi:2015,Klosta:2015,Hoover:2018,Schuech:2019,Nguyen:2019}. Often these tools require immense computational resources, e.g., high performance computing clusters, to simulate the system, either due to an individual simulation's computational time expense, the number of simulations necessary to generate data for a particular study, or both \cite{Borazjani:2012,Borazjani:2015,Bottom:2016,Hoover:2017,Hoover:2019,Miles:2019b}. Due to these restrictions, parameter sweeps are regularly performed with respect to one model parameter at a time, thus not fully mapping performance out across an entire parameter space, or even multi-dimension subspaces. $N$-dimensional parameter spaces are often studied through $1$-dimensional subspaces (lines) through the higher dimensional space \cite{Gutierrez:2014,Ngo:2014,Bale:2014,Hoover:2015,Montenegro:2016}. Thus, $2D$ (or $3D$) performance landscapes have not been fully mapped out, in which performance metrics can be interpreted from any combination of two (or more) traits.

Nematode locomotion has been well-studied for over half a century \cite{Gray:1964,Gray:1968,Jordan:1992,Gillis:1996,Ghosh:2008,Majmudar:2012,Padmanabhan:2012,Luersen:2014,Gutierrez:2014,Backholm:2015,Montenegro:2016}. Nematodes are an ideal model organism to study due to their simple two-dimensional planar gait. They display anguilliform modes of swimming; their muscles contract, resulting in their entire bodies bending, head to tail, with the locomotive benefit of propelling forward \cite{Gray:1964,Jordan:1992,Pierce-Shimomura:2008,Ghosh:2008,Backholm:2015}. Due to their bodylengths on the order of $\sim1$ mm \cite{Sznitman:2010}), many nematodes, like \textit{C. elegans} live at low Reynolds numbers ($\mathrm{Re}<1$). The Reynolds number, $\mathrm{Re}$, is non-dimensional quantity that is used to quantify the \textit{scale} of a fluid-system. It is defined in terms on four parameters: two system parameters - a characteristic length and velocity, $L$ and $V$, respectively, and two parameters describing physical properties of the underlying fluid - its density and dynamic viscosity, $\rho$ and $\mu$, respectively. The Reynolds number is defined as
\begin{equation}
    \label{eq:Re-intro} Re = \frac{\rho LV}{\mu}.
\end{equation}
Aquatic locomotion performance has been observed to be scale ($\mathrm{Re}$) dependent \cite{Purcell:1977,Smits:2019} in a number of organisms, such as water boatmen \cite{Ngo:2014}, fish \cite{Gazzola:2016}, and jellyfish \cite{Hershlag:2011,Miles:2019b}. As \textit{C. elegans} exist in the lower $\mathrm{Re}$ regime, many fluid-structure interaction computational models of nematodes have only considered the low Reynolds limit, i.e., the Stokes flow Regime ($\mathrm{Re}=0$) \cite{Berman:2013,Gutierrez:2014,Montenegro:2016}. However, anguilliform modes of locomotion are reliable swimming gaits for many organisms, such as eels or lamprey \cite{Tytell:2004,Hamlet:2015}, who perceive life through the lens of intermediate or higher Reynolds numbers, i.e., $\mathrm{Re}$ in the 100s, 1000s, or higher. Computational fluid dynamics models tend to be easier to perform fluid scaling studies than purely experimental studies, as they reduce the need for numerous high-fidelity scaled physical models and/or large quantities fluids of various viscosities, or finding organisms of specific sizes and/or training them. Furthermore, theoretical extensions of existing locomotion modes into different fluid scales are possible through computational modeling.

In this work, I attempt to quantify swimming performance using various metrics (forward swimming speed, an emergent peak-to-peak stroke amplitude, Strouhal number, cost of transport, a distance effectiveness ratio, and an angular trajectory metric), across broad $2$-dimensional parameter subspaces for an idealized, simple swimming model in $2D$ that resembles a nematode, like a \textit{C. elegans}. The simple swimmer model is a one-dimensional entity that propagates itself forward by dynamically changing its body's $1D$ preferred curvature state between two specific body positions - either a concave up or concave down state \cite{BattistaIB2d:2018,Battista:2020}, see Figure \ref{fig:Methods_Geo_Curvature}. Similar to a \textit{C. elegans}, the swimmer's locomotive patterns resemble a C-shape \cite{Jung:2010}, rather than a more S-shape, which \textit{C. elegans} use for crawling \cite{Pierce-Shimomura:2008} or like that of other angulliform swimmers like eels or lampreys \cite{Hamlet:2015}. The governing fluid-structure interaction equations governing this system are solved using the immersed boundary method (IB).

The parameter space explored is composed of the fluid scale (an \textit{input} Reynolds number, $\mathrm{Re}_{in}$), the stroke (undulation) frequency, $f$, and a kinematic parameter, $p$. The kinematic parameter, $p$, helps govern the kinematic profile of each stroke. Varying $p$ is akin to changing the acceleration and velocity of the undulation movement itself. That is, it controls how quickly each stroke accelerates from rest (current curvature state) to its maximal velocity and back to rest (next curvature state). Previous experimental work suggested that a nematode's undulatory amplitude did not vary when placed in a variety of increasing viscosity environments \cite{Korta:2007}, i.e., decreasing $\mathrm{Re}$; however, this may not be the case as $\mathrm{Re}$ increases, as this model demonstrates. The same nematode morphology was used across all simulations. However, previous work has established that bending frequency impacts nematode performance, for both speed and efficiency, at $\mathrm{Re}\sim0.4$ \cite{Luersen:2014}. Thus, I elected to vary stroke (undulation) frequency. Moreover, the swimming performance of C-shape undulation, like that of \textit{C. elegans}, has not been previously investigated across intermediate $\mathrm{Re}$.

Ultimately, I explored relationships between the three input parameters varied, $(\mathrm{Re}_{in},f,p)$, by analyzing two parameters at a time, while holding the third constant at three different values. Robust parameter subspaces were then identified, to which offered greater swimming performance than others. Moreover, from this data, Pareto-like fronts were uncovered for different parameter combinations \cite{Eloy:2013, Verma:2017,Smits:2019}.

%
%
%
%

\section{Methods}
\label{sec:methods}

\begin{figure}[H]
    \centering
    %
    \includegraphics[width=0.90\textwidth]{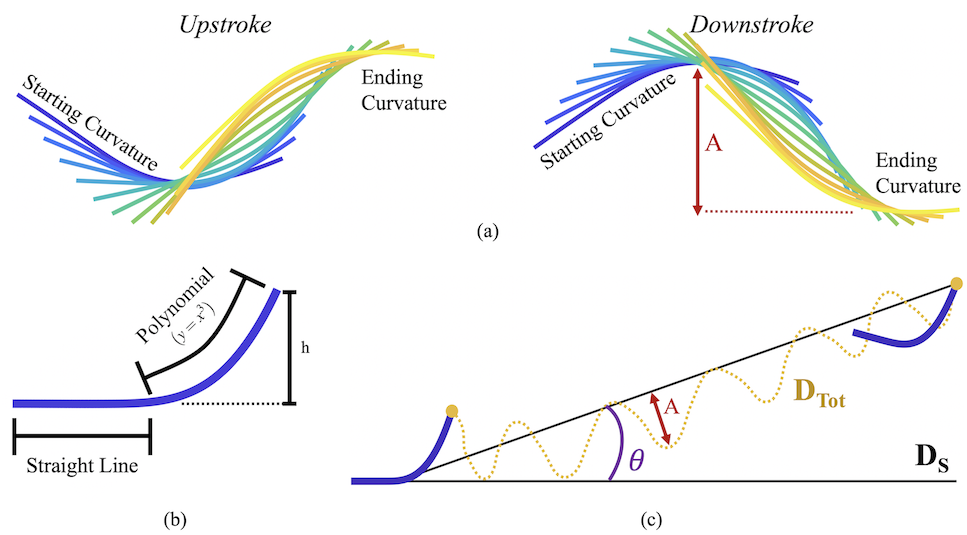}
    %
    %
    \caption{(a) Evolution of the preferred curvature during the upstroke and downstroke. The emergent peak-to-peak stroke amplitude, $A$, of each stroke is also depicted. (b) The geometry of the simple swimmer is composed of a straight line segment and a curve portion given by a cubic polynomial. (c) Depictions of the definitions for the angular trajectory as off the horizontal ($\theta$), the horizontal distance swam ($D_S$), and the total linear distance moved by the swimmer ($D_{Tot}$). Note that (c) depicts a swimmer for the case $(\mathrm{Re}_{in},f,p)=(1350,2.1,0.25)$.}
    \label{fig:Methods_Geo_Curvature}
\end{figure}

The idealized, simple swimming model first presented in \cite{BattistaIB2d:2018} and further explored in \cite{Battista:2020}, was modified in this work, see Figure \ref{fig:Methods_Geo_Curvature}b for its basic $1D$ geometry. This swimmer resembles a nematode; it is able to propagate forward by varying its curvature between two preferred curvature (shape) states as illustrated in Figure \ref{fig:Methods_Geo_Curvature}a - between a concave up state and a concave down geometric state. The choice of the body to be comprised of a straight section with a cubic-polynomial cap was to mitigate a simple asymmetry in the body itself, as observed in \textit{C. elegans} previously \cite{Ghosh:2008,Yuan:2015,Backholm:2015}. The straight portion of its body comprised 28\% of its entire bodylength ($L$), while the polynomial portion, composed the remaining 72\%. This choice was to give an initial height of $h=0.5$ (see Figure \ref{fig:Methods_Geo_Curvature}b). One complete stroke (undulation) was defined as comprising both the upstroke and downstroke of the swimmer's body itself.

To computationally model this swimmer, an open-source $2D$ implementation of the immersed boundary method \cite{Peskin:2002}, \textit{IB2d} \cite{Battista:2015,BattistaIB2d:2017}, was used. In fact, this swimmer is one of the built-in models within the software \cite{BattistaIB2d:2018}. It can be found at \url{github.com/nickabattista/IB2d} in the sub-directory: 
\begin{center}
\texttt{IB2d$/$matIB2d$/$Examples$/$Examples$\_$Education$/$Interpolation$/$Swimmer}.
\end{center}

The mathematical details regarding this swimmer's implementation in \textit{IB2d} can be found in the Supplemental Materials with further details in \cite{BattistaIB2d:2018,Battista:2020}. Additional details on the IB method can be found in \cite{Peskin:1996,Peskin:2002,Battista:2015,BattistaIB2d:2017}. Therefore I will only offer the computational parameters listed in Table \ref{table:num_param} here; however, the interpolation procedure for the swimmer's dynamical curvature is briefly described below, noting that the interpolant is detailed in greater depth in \cite{Battista:2020}. Each curvature state's information are contained in matrices \textbf{A} and \textbf{B}, for the concave up and concave down configurations, respectfully. Details regarding how the curvature is defined are found in the Supplemental Materials. I begin by defining a matrix function, $\textbf{h}(\tau)$, that interpolates the curvature of the swimmer between two \textit{curvature} states $\textbf{A}$ and $\textbf{B}$, 

\begin{equation}
    \label{cubic_interp} \textbf{h}(\tau) = \textbf{A} + g(\tau)( \textbf{B}-\textbf{A}),
\end{equation}
where $g(\tau)$ is a cubic polynomial defined to be
\begin{equation}
    \label{cubic_interp3} g(\tau) = \left\{\begin{array}{cc}
        g_0(t)=a_0 + a_1\tau + a_2\tau^2 + a_3\tau^3 \ \ & 0\leq \tau \leq p \\
        g_1(t)=b_0 + b_1\tau + b_2\tau^2 + b_3\tau^3 \ \ & p\leq \tau \leq 1-p \\
        g_2(t)=c_0 + c_1\tau + c_2\tau^2 + c_3\tau^3 \ \ & 1-p \leq \tau\leq 1
     \end{array}\right..
\end{equation}

\begin{figure}[H]
    \centering
    %
    \includegraphics[width=0.95\textwidth]{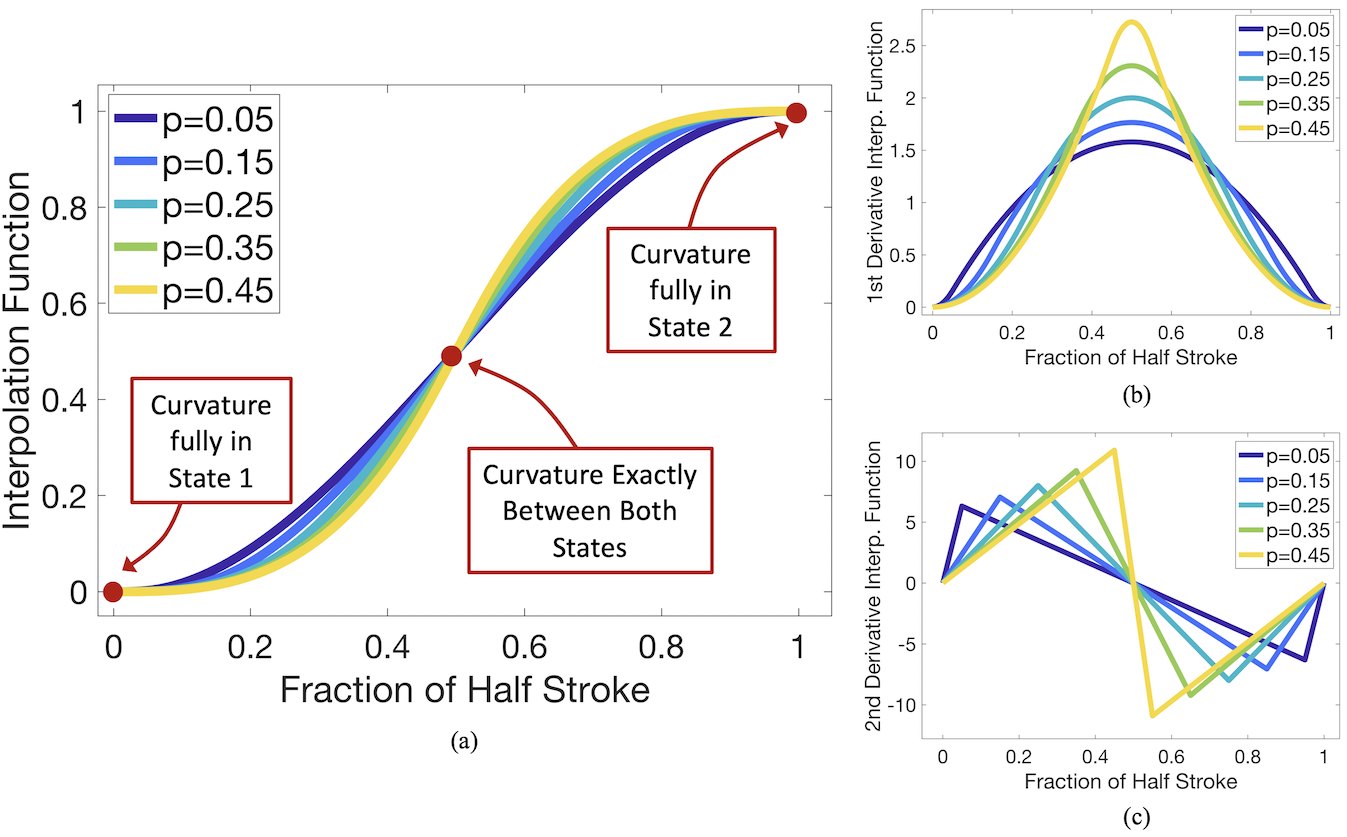}
    %
    %
    \caption{(a) The interpolation function, $g(\tau)$, (b) its first derivative, $g'(\tau)$, and (c) its second derivative, $g''(\tau)$, for a variety of $p$ values.}
    \label{fig:Methods_InterpFunction}
\end{figure}

$\tau$ is a non-dimensional time given by the fraction of either the upstroke or downstroke, i.e., a half-stroke. Since the period of a half-stroke is $0.5/f$, then $\tau = t/(0.5/f)$, as $0\leq t\leq 0.5/f$ and $0\leq\tau\leq1$ for each half-stroke. $p$ is an interpolation mediary point (the \textit{kinematic parameter}).  $p$ can be used to control the velocity and acceleration of how the curvature changes during the upstroke and downstroke, i.e., it controls the acceleration to the maximal speed in which the swimmer's curvature changes during either upstroke or downstroke, as well as the deceleration thereafter. Varying $p$ produces interpolants such as those in Figure \ref{fig:Methods_InterpFunction}. The the maximal and minimal acceleration occur at $p$ and $1-p$, respectively, see Figure \ref{fig:Methods_InterpFunction}c. The choice of a piecewise cubic interpolant (Eq. \ref{cubic_interp3}) was to ensure that there were enough degrees of freedom to allow for a kinematic control parameter ($p$) as well as to satisfy the following continuity conditions:

\begin{equation}
    \left.\begin{array}{c} \textbf{h}(0) = \textbf{A} \\  \textbf{h}(1) = \textbf{B}  \end{array}\right\} \ \mbox{continuity} \ \ \
    \left.\begin{array}{c} \textbf{h}'(0) = 0 \\  \textbf{h}'(1) = 0  \end{array}\right\} \ \mbox{cont. velocities} \ \ \
    \left.\begin{array}{c} \textbf{h}''(0) = 0 \\  \textbf{h}''(1) = 0  \end{array}\right\} \ \mbox{no instant. accel.}
\end{equation}

\begin{table}
\begin{center}
\begin{tabular}{| c | c | c | c |}
    \hline
    Parameter               & Variable    & Units        & Value \\ \hline
    Domain Size            & $[L_x,L_y]$  & m               &  $[6,16]$             \\ \hline
    Spatial Grid Size      & $dx=dy$      & m               &  $L_x/1024=L_y/384$            \\ \hline
    Lagrangian Grid Size    & $ds$        & m               &  $dx/2$               \\ \hline
    Time Step Size          & $dt$        & s               &  $2.5\times10^{-5}$   \\ \hline
    Total Simulation Time    & $T$        & \textit{stroke cycles} &  $6$               \\ \hline
    Fluid Density            & $\rho$     & $kg/m^3$        &  $1000$               \\ \hline
    Fluid Dynamic Viscosity & $\mu$      & $kg/(ms)$       &  [0.475,17750]       \\ \hline
    Swimmer Length           & $L$        & m               &  $1.42$    \\ \hline
    Swimmer Height           & $h$        & m               &  $0.5$        \\ \hline
    Stroke Frequency         & $f$        & $s^{-1}$       &   [1,2.5]  \\ \hline
    \textit{Input} Reynolds number         & $\mathrm{Re}_{in}$        & -              &  [0.1,4500] \\ \hline
    Kinematic Parameter     & $p$        & $-$            &  [0.075,0.425] \\ \hline
    Spring Stiffness   & $k_{spr}$ & $kg\cdot m/s^2$ &  $9.5625\times10^{9}$  \\ \hline
    Non-invariant Beam Stiffness   & $k_{beam} $ & $kg\cdot m/s^2$ &  $2.03634\times10^{12}$  \\ \hline
    \end{tabular}
    \caption{Numerical parameters used in the two-dimensional simulations.}
    \label{table:num_param}
    \end{center}
\end{table}

In this study, three parameters were varied - the dynamic viscosity ($\mu$), the stroke frequency ($f$), and a kinematic parameter, $p$. The viscosity $\mu$ was varied to effectively change an input fluid scale, given by an \textit{input} Reynolds number, $\mathrm{Re}_{in}$, i.e.,
\begin{equation}
    \label{eq:Re} \mathrm{Re}_{in} = \frac{\rho L (fL)}{\mu}. 
\end{equation}
The characteristic length and velocity scale are defined to be the swimmer's bodylength, $L$, and a frequency-based velocity, given by $fL$ \cite{Cui:2017,Dai:2018}. For each simulation a specific $(\mathrm{Re}_{in},f,p)$ combination was determined, to which the corresponding $\mu$ was computed from Eq. \ref{eq:Re} to give the appropriate $\mathrm{Re}_{in}$. Note that it is customary in anguilliform studies to use $fA$, the product of $f$ and the peak-to-peak undulation amplitude, $A$, as the characteristic velocity in the Reynolds number calculation. However, the undulation amplitude is an emergent output of the model and thus cannot be known when initializing the simulations. Therefore I elected to use a characteristic velocity seen in fish literature as an input velocity scale, $fL$ \cite{Cui:2017,Dai:2018,Cui:2020} and show relationships between an input Reynolds number ($\mathrm{Re}_{in}$) and output Reynolds number ($\mathrm{Re}_{out}$). The output Reynolds number is defined as $\mathrm{Re}_{out}=\rho\cdot L\cdot fA/\mu$, whose frequency based velocity scale is given as $fA$. Relationships among $\mathrm{Re}_{in}$ and $\mathrm{Re}_{out}$ are depicted in Figure S1 in the Supplemental Materials. The output Reynolds numbers fall within the intermediate $\mathrm{Re}$ regime, which is an interesting regime to study due to the balance of inertial and viscous forces \cite{Klotsa:2019}. Moreover, $\mathrm{Re}_{out}$ appears most strongly correlated to $\mathrm{Re}_{in}$, compared to either $f$ or $p$. Also, the definition of the output peak-to-peak stroke amplitude, $A$, is illustrated in Figure \ref{fig:Methods_Geo_Curvature}a and c.

The model output included a non-dimensional forward swimming speed, which was defined to be the inverse of the Strouhal number, i.e.,
\begin{equation}
    \label{eq:Strouhal} \mathrm{St} = \frac{fA}{V_F},
\end{equation}
where $\mathrm{St}$ is the Strouhal number, $V_F$ is the dimensional forward swimming speed, and $fA$ is a frequency based velocity based on the \textit{output} peak-to-peak stroke amplitude of the swimmer, $A$. The above definition of $\mathrm{St}$ uses the stroke (undulation) frequency, $f$, rather than a vortex shedding frequency, as is common in swimming studies. The $\mathrm{St}$ analysis focused on the $\mathrm{St}$ range in which the majority of swimming and flying animals reside, i.e., $0.2<\mathrm{St}<0.4$ \cite{Taylor:2003}. Moreover, a power-based cost of transport \cite{Bale:2014,Hamlet:2015} was computed and defined to be

\begin{equation}
    \label{eq:COT} COT_{Dim} = \frac{1}{N} \frac{1}{V_F} \displaystyle\sum_{j=1}^N |F_j||U_j|,
\end{equation}
where $F_j$ and $U_{j}$ are the applied vertical force and tangential body velocity of the swimmer at $N$ time-points during a specific period of time. Note that since $dt$ is fixed and the frequency changes across many simulations, the number of time-steps may vary from simulation to simulation. However, each simulation's time-steps were sub-sampled at specific time-points of each stroke cycle to ensure that data was stored at the same fractions of a stroke cycle across all simulations performed. These differences were taken into account when computing all locomotion quantities used in the performance metrics, i.e., $V_F$, $F_j$, and $U_j$. The cost of transport was non-dimensionalized, in the following manner $$COT = \frac{COT_{Dim}}{\rho f^{2} A^2 L^2},$$ where $L$ is the length of the swimmer. The non-dimensional $COT$ is similar to finding the energy-consumption coefficient of \cite{Bale:2014}. Also, a distance effectiveness metric was computed, $d_{eff}$. It was defined as the ratio of forward distance swam and the total linear distance swam, i.e.,
\begin{equation}
    \label{eq:DistEff} d_{eff} = \frac{D_S}{D_{Tot}},
\end{equation}
where $D_S$ and $D_{Tot}$ are the horizontal (forward) distance and the total linear distance moved by the swimmer, respectively, during the same $N$ time-points, see Figure \ref{fig:Methods_Geo_Curvature}c. $D_{Tot}$ was computed by tracking the linear distance traveled by the swimmer's head across the $N$ time-points and adding them together. Thus, $D_{Tot}$ encompassed both vertical and horizontal movement. Lastly, the average angle off the horizontal, $\theta$, was computed (see Figure \ref{fig:Methods_Geo_Curvature}c), to discern parameter combinations that lead to non-horizontal swimming trajectories.

For each simulation performed, a time-averaged non-dimensional forward swimming speed ($1/\mathrm{St}$), Strouhal number ($\mathrm{St}$), $COT$, and peak-to-peak stroke amplitude ($A/L$) were computed along with $\theta$ and the distance effectiveness ratio, $d_{eff}$, which allowed for effectively mapping these metrics across entire broad subspaces when a sufficient number of simulations were performed.

%
%
%
%

\section{Results}
\label{results}

Different modes of locomotion are more effective at certain fluid scales than others \cite{Vogel:1996}. To investigate the effectiveness of this nematode-like, anguilliform swimming modality, simulations were performed across four orders of magnitude of $\mathrm{Re}_{in}$. Moreover, a plethora of literature has illustrated that an organism's forward swimming speed is dependent on its stroke (undulation) frequency \cite{Bainbridge:1958,Gray:1968,Steinhausen:2005}. By highly resolving a subspace of $(\mathrm{Re}_{in},f)$ allowed for exploring any nonlinear effects that could arise when both $\mathrm{Re}_{in}$ and $f$ are varied for this particular mode of locomotion. Furthermore, by varying the kinematic parameter $p$, one could explore how any such nonlinear affects might be exacerbated by variations of the velocity/acceleration within the stroke's (undulation's) kinematic profile itself. Thus, a total of 6,357 fully coupled 2D fluid-structure interaction simulations were performed to explore the $3$-dimensional parameter space for this idealized, simple swimming model. With the goal of finding robust parameter subspaces which lead to higher swimming performance, the following cases were considered:
\begin{enumerate}
    \item The \textit{input} Reynolds number \& frequency space for $3$ specific values of the kinematic parameter, $p$.
    \item The \textit{input} Reynolds number  \& kinematic parameter space for $3$ specific frequencies.
    \item The frequency \& kinematic parameter space for $3$ specific \textit{input} Reynolds numbers
\end{enumerate}

\begin{figure}[H]
    \centering
    %
    \includegraphics[width=0.975\textwidth]{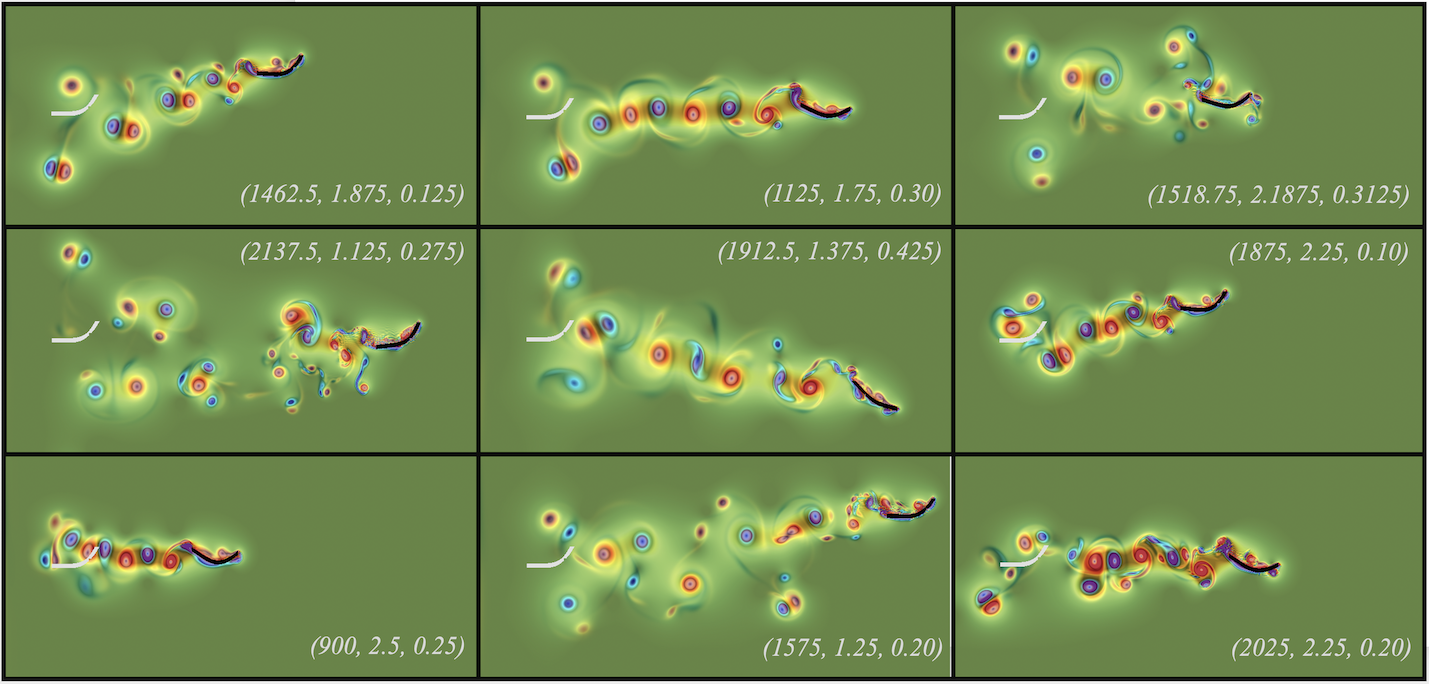}
    %
    %
    \caption{Numerous swimmer's vortex wake and position after its $5^{th}$ full stroke cycle. Different parameter combinations lead to different swimming behavior, both in terms of distance swam and dynamics, as seen by variations in the vortex wakes left behind. The colormap illustrates vorticity and the initial position of the swimmer is given for each case.}
    \label{fig:VorticityComparison}
\end{figure}

Varying these three parameters $(\mathrm{Re}_{in},f,p)$ can lead to substantially different swimming behavior, both in terms of kinematics as well as dynamical performance. Figure \ref{fig:VorticityComparison} gives snapshots of the swimmer's position and fluid vorticity after 5 complete stroke cycles (5 upstrokes and 5 downstrokes each) for a variety of  $(\mathrm{Re}_{in},f,p)$ combinations. As parameters are varied, some swimmers are able to outperform others in terms of distance swam and forward speed; however, not all swimmers move laterally across. Some swimmers begin drifting upwards or downwards as well at different angular trajectories.

%
%

\subsection{Exploring the Reynolds number \& Frequency Space}
\label{results:ReFreq}

\begin{figure}[H]
    \centering
    %
    \includegraphics[width=0.90\textwidth]{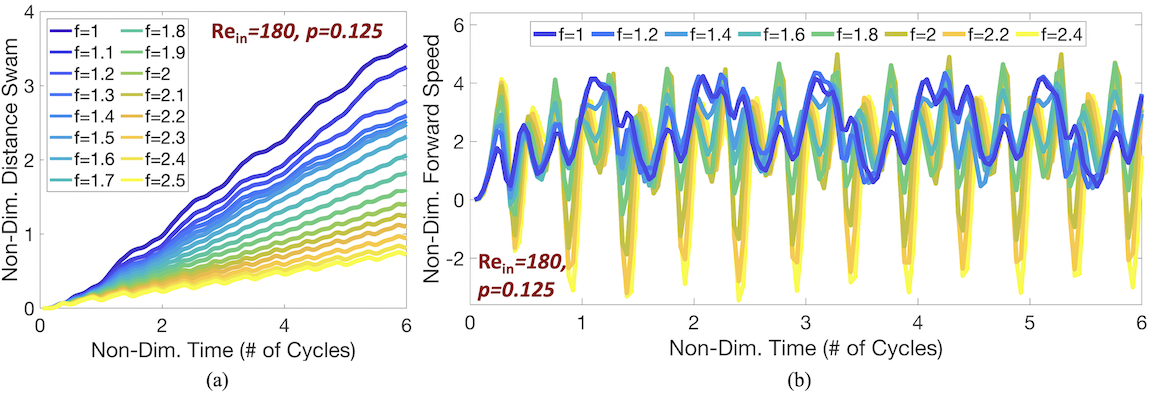}
    %
    %
    \caption{The (a) non-dimensional distance (bodylengths)  and (b) speed ($1/\mathrm{St})$ over time for swimmers with $\mathrm{Re}_{in}=180$, $p=0.125$ and over variety of $f$.}
    \label{fig:ReVsFreq_Re20_Data}
\end{figure}

First, an input fluid scale ($\mathrm{Re}_{in}$) and stroke frequency ($f$) subspace was explored for $3$ values of the kinematic parameter, $p$. Generally, as $\mathrm{Re}_{in}$ increased, the swimming speed $(1/\mathrm{St})$ increased for a given $f$. Lower $f$ tended to have higher swimming speeds for a given $\mathrm{Re}_{in}$. Figure \ref{fig:ReVsFreq_Re20_Data} illustrates that both the distance swam and swimming speed were greater for lower frequencies for the case of $\mathrm{Re}_{in}=180$. Moreover, the swimming speed's waveform changes as frequency changes. Holding a fixed $f$ and varying $Re$ produces similar trends, but where higher $\mathrm{Re}_{in}$ leads to enhanced swimming. For this data, see Figure S2 in the Supplemental Documents. 

\begin{figure}[H]
    \centering
    %
    \includegraphics[width=0.825\textwidth]{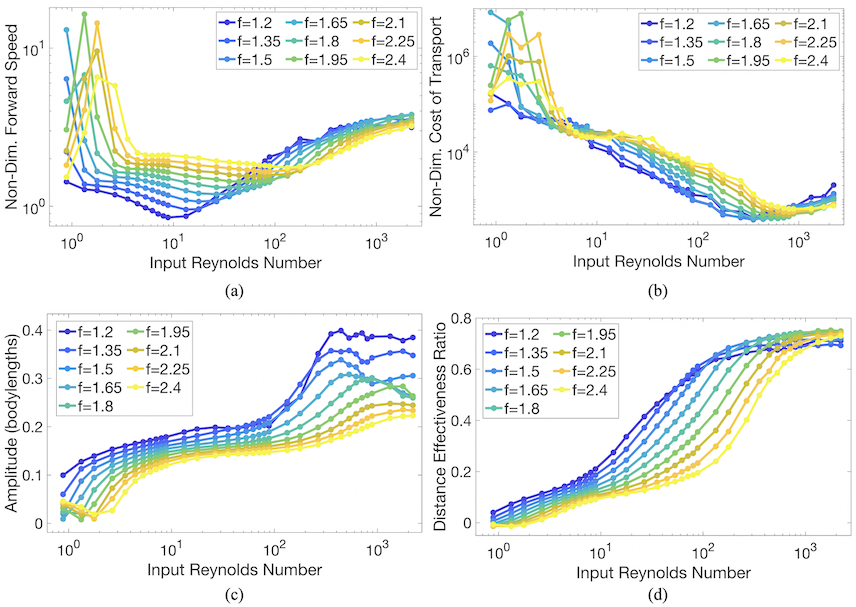}
    %
    %
    \caption{The time-averaged data on the $(\mathrm{Re}_{in},f)$-subspace for $p=0.125$ for (a) non-dimensional swimming speeds ($1/\mathrm{St}$), (b) non-dimensional cost of transports ($COT$), (c) peak-to-peak output stroke amplitudes ($A$) , and (d) non-dimensional distance effectiveness metrics ($d_{eff}$).}
    \label{fig:ReVsFreq_4Data_Plot}
\end{figure}

However, nonlinear relationships exist between swimming speed, $\mathrm{Re}_{in}$, and $f$, for given $p$. Figure \ref{fig:ReVsFreq_4Data_Plot} provides temporally-averaged data for forward swimming speed, $\mathrm{St}$, cost of transport ($COT$), peak-to-peak stroke amplitude ($A/L$), and $d_{eff}$ for a subset of the total cases performed involving differing $\mathrm{Re}_{in}$ and $f$ for $p=0.125$. Swimming speed $(1/\mathrm{St})$ and cost of transport ($COT$) take on maximal value between $1\lesssim \mathrm{Re}_{in}\lesssim5$; however, this is due to very small emergent stroke amplitudes. For $5\lesssim \mathrm{Re}_{in}\lesssim 100$, higher $f$ produces faster swimmers; however, near $\mathrm{Re}_{in}=100$, a transition occurs in which lower $f$ produces faster swimmers. Throughout the range of $5\lesssim \mathrm{Re}_{in}\lesssim 500$, $COT$ decreases; it begins to increase shortly after $\mathrm{Re}_{in}\approx500$.

\begin{figure}[H]
    \centering
    %
    \includegraphics[width=0.90\textwidth]{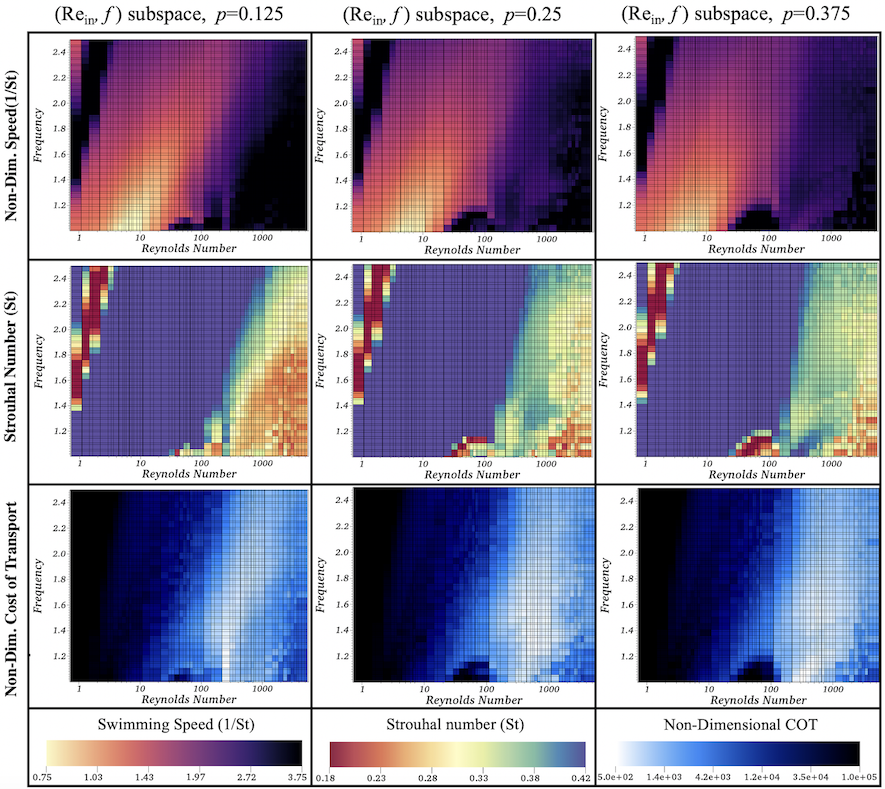}
    %
    %
    \caption{Colormaps illustrating the non-dimensional forward swimming speeds (top row) and cost of transports (bottom row) over the $(\mathrm{Re}_{in},f)$-subspaces for $3$ different $p$ values: $p=0.125,0.25,$ and $0.375$.}
    \label{fig:ReVsFreq_3P_Colormaps}
\end{figure}

For comparative purposes, Figure \ref{fig:ReVsFreq_3P_Colormaps} presents the swimming speed $(1/\mathrm{St})$, $\mathrm{St}$, and $COT$ data as colormaps for all cases of $p$ considered ($p=0.125,0.25,$ and $0.375$). Each shaded-in box in the grid corresponds to an individual, independent FSI simulation of a different $(\mathrm{Re}_{in},f)$ combination for a specific $p$. Generally a combination of a higher $\mathrm{Re}_{in}$ ($\mathrm{Re}_{in}\gtrsim300$) and lower $f$ produces the fastest swimmers; however, the faster swimmers also tend to correspond to the slightly more costly swimming, i.e., the non-minimal regions of $COT$ are near regions of maximal swimming speed, and higher stroke amplitudes. Lower $f$ resulted in higher stroke amplitudes in general. As $p$ varies, there appear to be only subtle differences in these performance metrics across the $(\mathrm{Re}_{in},f)$-subspace. The region of $0.2<\mathrm{St}<0.4$ as well as higher $d_{eff}$ slightly decreases in each subspace as $p$ increases (see Figure S3 in the Supplemental Materials). Furthermore, at low $\mathrm{Re}_{in}$ the swimmer appears to substantially drift downward (negative $\theta$) while swimming. As $\mathrm{Re}_{in}$ increases there is an abrupt transition to the swimmer migrating upwards (positive $\theta$). For $\mathrm{Re}_{in}\gtrsim 30$ the swimmer has less of a vertical shift off the horizontal as it swims (see Figure S3). The dimensional analog to Figure \ref{fig:ReVsFreq_3P_Colormaps} is also provided as Figure S17 in the Supplemental Materials.

However, Figure \ref{fig:ReVsFreq_pSens_3Plot} presents the swimming speeds observed for all frequencies considered and three different kinematic parameters for 3 different $\mathrm{Re}_{in}$ cases: (a) $\mathrm{Re}_{in}=6$, (b) $\mathrm{Re}_{in}=45$, and (c) $\mathrm{Re}_{in}=270$. For a given $Re$, swimming speed appears more sensitive to frequency and less sensitive to variations in $p$, e.g., in the case of $\mathrm{Re}_{in}=6$, varying $f$ could result in $\sim3$-fold difference in swimming speeds, while varying $p$ did not seem to produce a substantial difference. A figure giving the swimming speed data in different dimensional forms for the $p=0.125$ case is provided in Figure S4 in the Supplemental Materials. For $\mathrm{Re}_{in}\gtrsim 100\ \mathrm{Hz}$, Figure S4c shows a linear relationship between swimming speed (bodylengths/stroke) and the logarithm of $\mathrm{Re}_{in}$.

\begin{figure}[H]
    \centering
    %
    \includegraphics[width=0.975\textwidth]{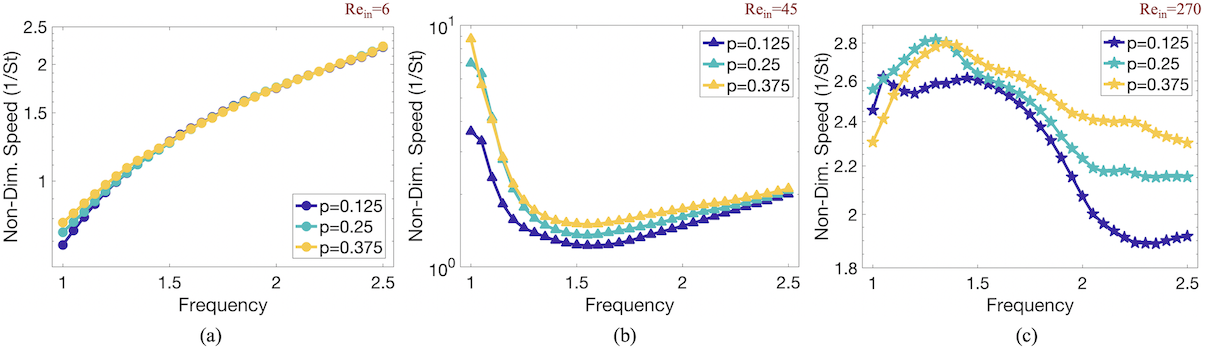}
    %
    %
    \caption{Semi-logarithmic plots illustrating the sensitivity of forward swimming speed to $p$ for particular cases of $Re$: (a) $\mathrm{Re}_{in}=6$, (b) $\mathrm{Re}_{in}=45$, and (c) $\mathrm{Re}_{in}=270$.}
    \label{fig:ReVsFreq_pSens_3Plot}
\end{figure}

%
%

\subsection{Exploring the input Reynolds number \& Kinematic Parameter Space}
\label{results:Re_pVal}

\begin{figure}[H]
    \centering
    %
    \includegraphics[width=0.90\textwidth]{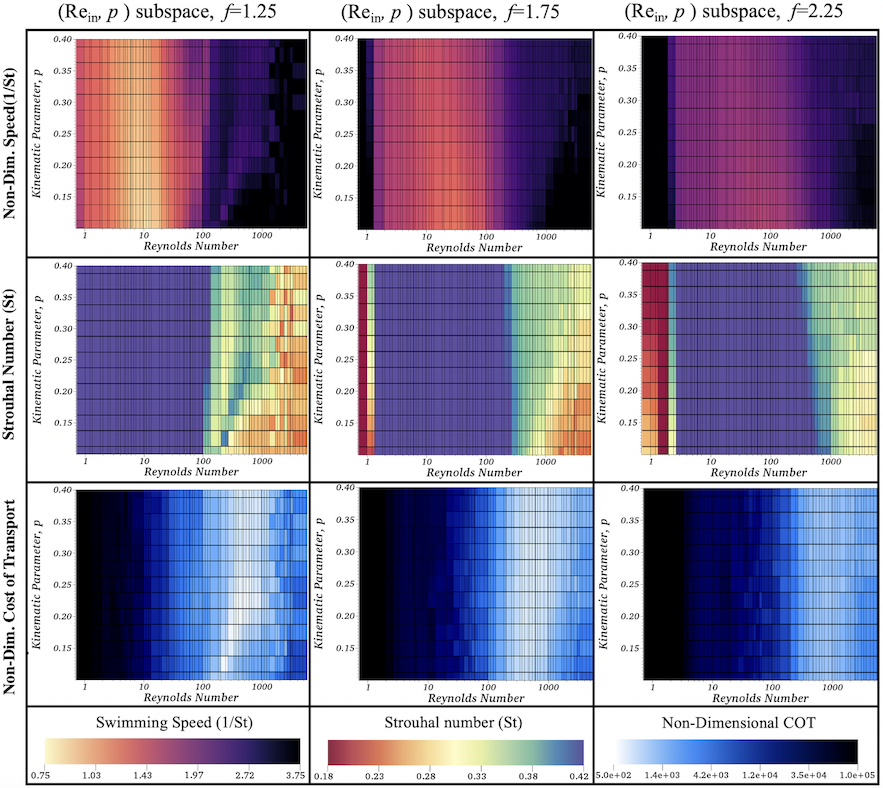}
    %
    %
    \caption{Colormaps illustrating the non-dimensional forward swimming speeds (top row) and cost of transports (bottom row) over the $(\mathrm{Re}_{in},p)$-subspaces for $3$ different $f$ values: $f=1.25, 1.75$ and $2.25\ \mathrm{Hz}$.}
    \label{fig:ReVsP_SpeedCOT_Colormaps}
\end{figure}

The $(\mathrm{Re}_{in},p)$-subspace was explored for three particular $f$ values: $f=1.25, 1.75,$ and $2.25\ \mathrm{Hz}$. As suggested by Figure \ref{fig:ReVsFreq_3P_Colormaps}, varying $p$ for a given $\mathrm{Re}_{in}$, leads to smaller variations in output metrics. For the cases of $\mathrm{Re}_{in}=180$ and $f=1.75$ and $2\ \mathrm{Hz}$, the distance swam and swimming speed over time are given in Figures S6 and S7 in the Supplemental Materials. Moreover, as $p$ is varied for $\mathrm{Re}_{in}\lesssim500$ (and particular $f$), there are minimal differences in forward swimming speed and $COT$, see Figure \ref{fig:ReVsP_SpeedCOT_Colormaps}. Thus overall the time-averaged performance metrics, such as forward speed $(1/\mathrm{St})$ and $COT$, do not appear too sensitive to variations in $p$ for $\mathrm{Re}_{in}\lesssim 500$. However, $\mathrm{Re}_{in}\gtrsim500$ and lower $f$ correspond larger regions of maximal swimming, but with smaller stroke amplitudes (see Figure S9). Moreover, generally as $f$ increases, stroke amplitude decreases for $\mathrm{Re}_{in}\gtrsim1$. Note that the raw data can be seen in Figure S8 provided in the Supplemental Materials. Similar trends can be seen for $d_{eff}$ and $\theta$, see Figure S9 in the Supplemental Materials.

However, Figures \ref{fig:ReVsP_SpeedCOT_Colormaps} and S9 illustrate that as $f$ varies, the overall performance depicted on each $(\mathrm{Re}_{in},p)$-subspace greatly varies. Higher $\mathrm{Re}_{in}$ generally led to greater swimming speeds, $\mathrm{St}$ values that fell within the biologically relevant range ($0.2<St<0.4$), as well as higher $d_{eff}$, suggesting that the movements of the swimmer led to increased forward thrust resulting in greater forward propagation. Higher $\mathrm{Re}_{in}$ also appears to result less upwards migration of the swimmer across the domain, i.e, the angle off the horizontal, $\theta$, is closer to 0. Although, for higher $f$ and low $\mathrm{Re}_{in}$, the subspace of drifting vertically downwards ($\theta\sim-25\degree$) continues to increase in size, followed by a regime of $\mathrm{Re}_{in}$ to which the swimmer drifts upwards ($\theta\sim30\degree$). Furthermore, a nonlinear relationship emerges in $COT$, where $COT$ takes on its minimal values over a subset of $\mathrm{Re}_{in}$ ($100\lesssim \mathrm{Re}_{in}\lesssim 1000$), with increasing values of $COT$ on either side of that subset. The dimensional analog to Figure \ref{fig:ReVsP_SpeedCOT_Colormaps} is provided as Figure S18 in the Supplemental Materials.

Figure \ref{fig:ReVsP_fSens_3Plot} shows that as $p$ varies for multiple $f$ and $\mathrm{Re}_{in}$: (a) $\mathrm{Re}_{in}=6$, (b) $\mathrm{Re}_{in}=45$, and (c) $\mathrm{Re}_{in}=270$, that changes in $p$ could vary forward swimming speeds by upwards of $\sim10-50\%$. This occurs in both the case of $\mathrm{Re}_{in}=45$ and $\mathrm{Re}_{in}=270$ for all $f$ considered. However, the $\mathrm{Re}_{in}=6$ case illustrates that varying $f$ results in two-fold increases ($\sim200\%$) in swimming speed, while varying $p$ minimally affects swimming speed. Moreover, increases in swimming speed by upwards of $\sim40\%$ can be seen for other values of $f$ in the $\mathrm{Re}_{in}=45$ and $270$ cases. Hence varying both $f$ an $p$ can greatly affect the swimmer's resulting swimming performance, depending on the fluid scale being explored, although variations in $f$ seems to more substantially affect the model. A figure giving the swimming speed data in different dimensional forms for the $f=1.75\ \mathrm{Hz}$ case is provided in Figure S10 in the Supplemental Materials. Only minute increases in swimming speed occur as $\mathrm{Re}_{in}$ increases from 0.5 to 100 for swimming speeds in units of bodylengths/stroke or bodylengths/second; however, shortly after $\mathrm{Re}_{in}>100$, speeds substantially increase.

\begin{figure}[H]
    \centering
    %
    \includegraphics[width=0.975\textwidth]{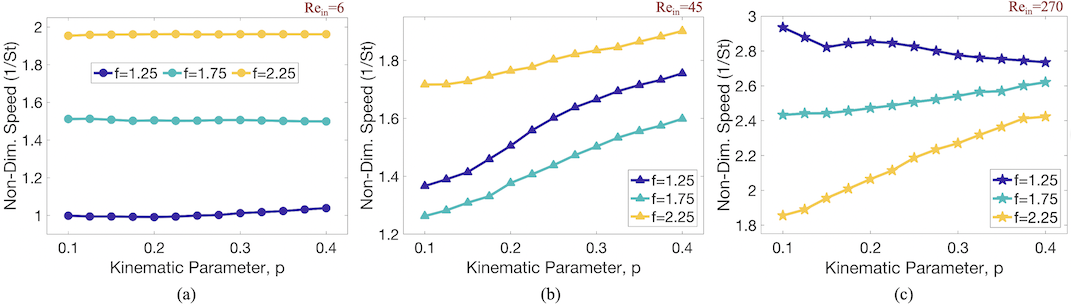}
    %
    %
    \caption{Plots illustrating the sensitivity of forward swimming speed to $f$ for particular cases of $\mathrm{Re}_{in}$: (a) $\mathrm{Re}_{in}=6$, (b) $\mathrm{Re}_{in}=45$, and (c) $\mathrm{Re}_{in}=270$.}
    \label{fig:ReVsP_fSens_3Plot}
\end{figure}

%
%

\subsection{Exploring the Frequency \& Kinematic Parameter Space}
\label{results:Freq_pVal}

\begin{figure}[H]
    \centering
    %
    \includegraphics[width=0.90\textwidth]{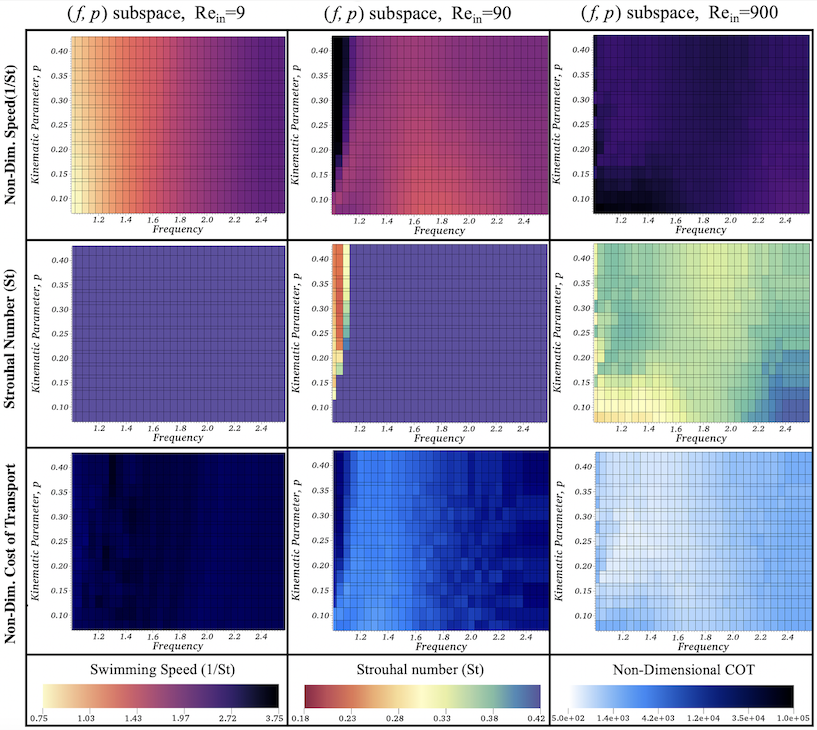}
    %
    %
    \caption{Colormaps illustrating the non-dimensional forward swimming speeds ($1/\mathrm{St}$) (top row) and cost of transports (bottom row) over the $(f,p)$-subspaces for $3$ different $\mathrm{Re}_{in}$ values: $\mathrm{Re}_{in}=9, 90$ and $900$.}
    \label{fig:FreqVsP_SpeedCOT_Colormaps}
\end{figure}

Lastly, the frequency and kinematic parameter subspace, $(f,p)$, was explored for $3$ specific $\mathrm{Re}_{in}$, $\mathrm{Re}_{in}=9, 90,$ and $900$. For $\mathrm{Re}_{in}=90$ minimal differences were observed in the distance swam and swimming speed $(1/\mathrm{St})$ over time (see Figure S12 in the Supplemental Materials). However, as $f$ changes substantial variations in swimming performance were observed, as shown in Figure \ref{fig:FreqVsP_SpeedCOT_Colormaps}, which illustrates the swimming speed $(1/\mathrm{St})$ and $COT$ on 3 $(f,p)$-subspaces corresponding to different $\mathrm{Re}_{in}$. In the $\mathrm{Re}_{in}=900$ case, the highest speeds were associated with low $f$ and low $p$. However, frequencies within the range of $1.6-2.0\ \mathrm{Hz}$ with $p>0.15$ appear to be associated with higher swimming speeds, than those of $f$ outside that range. This region also had generally lower $COT$. On the other hand, in the $\mathrm{Re}_{in}=9$ case, as $f$ increased, swimming speed generally increased while $COT$ generally decreased. The case in-between ($\mathrm{Re}_{in}=90$) shows the regions with the highest swimming speeds correpond to regions of the highest $COT$. However, substantial swimming speeds with low $COT$ are observed for lower $f$ and lower $p$. Similar speeds are also produced at higher $f$ and higher $p$ but with higher associated $COT$. Hence there exists a $\mathrm{Re}_{in}$ regime in which lower frequency results in higher swimming speed and lower $COT$. 

\begin{figure}[H]
    \centering
    %
    \includegraphics[width=0.975\textwidth]{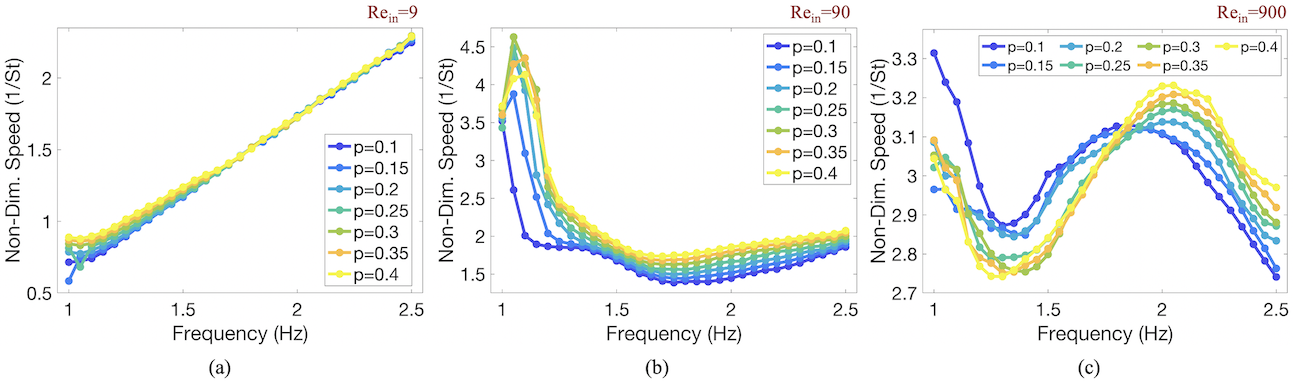}
    %
    %
    \caption{Plots illustrating how sensitive forward swimming speed is to $p$ for particular cases of $\mathrm{Re}_{in}$: (a) $\mathrm{Re}_{in}=9$, (b) $\mathrm{Re}_{in}=90$, and (c) $\mathrm{Re}_{in}=900$.}
    \label{fig:FreqVsP_ReSens_3Plot}
\end{figure}

There are intricate nonlinear relationships between swimming speed and $COT$ in certain $(f,p)$ subspaces that correspond to different $\mathrm{Re}_{in}$. Furthermore, these nonlinear relationships can be seen in Figure \ref{fig:FreqVsP_ReSens_3Plot}, which provides the time-averaged swimming speed ($1/\mathrm{St})$ for all frequencies considered and a subset of $p$ for three $\mathrm{Re}_{in}$: (a) $\mathrm{Re}_{in}=9$, (b) $\mathrm{Re}_{in}=90$, and (c) $\mathrm{Re}_{in}=900$. As $\mathrm{Re}_{in}$ increases from (a)$\rightarrow$(c), the shape of the curve changes. For $\mathrm{Re}_{in}=9$, increasing frequency resulted in monotonically  increasing forward swimming speeds, as previously suggested in Figure \ref{fig:ReVsFreq_4Data_Plot}. While for $\mathrm{Re}_{in}=90$, as frequency increased, there was a sudden drop in swimming speeds, followed by speeds increasing slightly for $f\gtrsim 1.75\ \mathrm{Hz}$. However, when $\mathrm{Re}_{in}=900$, a clear nonlinear relationship emerges, where a unique maxima and minima emerge near $f\sim1.3$ and $2 \mathrm{Hz}$, respectively.
Additional data for $1/\mathrm{St}$ $\mathrm{St}$, $A/L$, and $COT$ for the $\mathrm{Re}_{in}=90$ case can be found in Figure S13 in the Supplemental Materials. Similarly, colormaps of $d_{eff}$, $\theta$, and amplitude ($A/L$) can be found in Figure S14 in the Supplemental Materials. Note that swimming in the $0.2<\mathrm{St}<0.4$ range does not occur until the highest $\mathrm{Re}_{in}$ case considered ($\mathrm{Re}_{in}=900$). Also as $\mathrm{Re}_{in}$ increases, the $d_{eff}$ increases as well, indicating that the movement patterns the swimmer performs, leads to greater thrusts forward for swimming. This also tends to correspond to less vertical drifting when swimming, i.e., $\theta$ is closer to zero degrees.

Moreover, among all cases shown in Figures \ref{fig:FreqVsP_SpeedCOT_Colormaps} and \ref{fig:FreqVsP_ReSens_3Plot}, varying $p$ did not generally appear to substantially affect either forward speed or $COT$, i.e., swimming performance appears to be more sensitive to $f$ than to $p$. As Figure \ref{fig:FreqVsP_ReSens_3Plot}b illustrates, only in the higher $\mathrm{Re}_{in}$ cases of $\mathrm{Re}_{in}=90$ and $900$ did swimming performance appear sensitive to $p$, i.e., swimming speed is 1.3x larger in the $p=0.4$ case when compared to the $p=0.1$ case for $\mathrm{Re}_{in}=90$ and $f\approx 1.75\ \mathrm{Hz}$. A figure giving the swimming speed data in different dimensional forms for the $\mathrm{Re}_{in}=90$ case is provided in Figure S15 in the Supplemental Materials. For $f\gtrsim 1.75\ \mathrm{Hz}$, Figure S15c shows a linear relationship between swimming speed (bodylengths/second) and stroke frequency, while swimming speed (bodylengths/stroke) is constant in the same frequency range, see Figure 15b.

%
%

\subsection{Exploring Cost of Transport vs. Forward Swimming Speed}
\label{results:PhasePlane}

\begin{figure}[H]
    \centering
    %
    \includegraphics[width=0.85\textwidth]{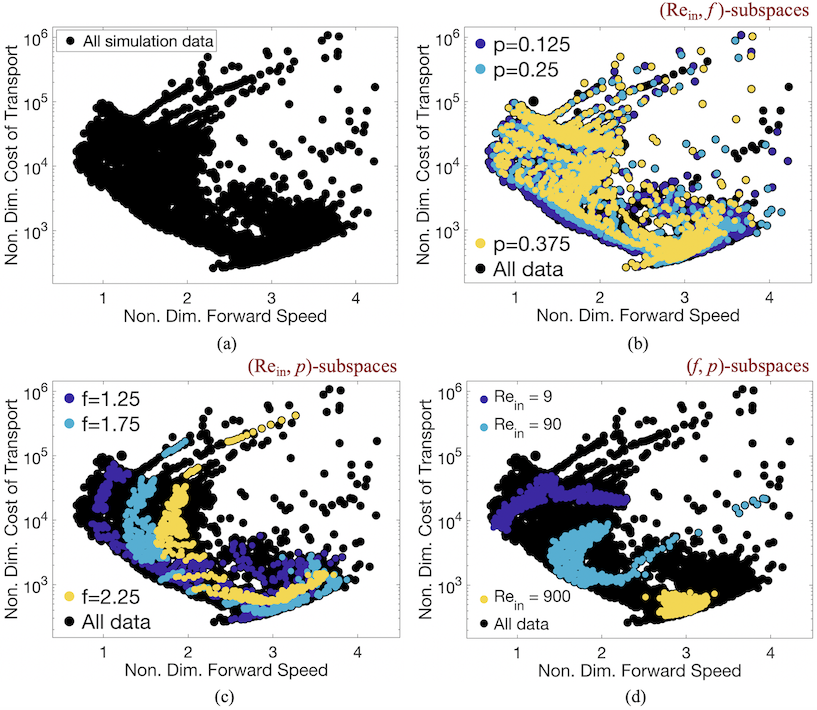}
    %
    %
    %
    \caption{The non-dimensional $COT$ and forward swimming speeds ($1/\mathrm{St}$) plotted against each other for (a) every simulation performed, (b) the $(\mathrm{Re}_{in},f)$-subspace, (c) the $(\mathrm{Re}_{in},p)$-subspace, and (d) the $(f,p)$-subspace.}
    \label{fig:Results_COT_vs_Speed}
\end{figure}

A Pareto-like front can be observed when plotting the non-dimensional cost of transport ($COT$) against the non-dimensional swimming speed ($1/\mathrm{St}$) \cite{Eloy:2013,Verma:2017,Schuech:2019,Smits:2019}, see Figure \ref{fig:Results_COT_vs_Speed}a. This is called \textit{Pareto-like} because maximal swimming speeds are desired for low $COT$, i.e., ones ideally hopes to minimize $COT$, rather than maximize it, and hence not maximize both quantities, as is traditionally done in Pareto optimization strategies. $COT$ appears to be minimal in cases when the swimmer's forward swimming speed is $2.5\lesssim 1/\mathrm{St} \lesssim 3$. The further outside of that range, the more $COT$ appears to increase. As Figure \ref{fig:Results_COT_vs_Speed}a presents the data from Sections \ref{results:ReFreq}-\ref{results:Freq_pVal} all combined, patterns emerge of where subspaces lie within the overall performance space, i.e., Figures \ref{fig:Results_COT_vs_Speed}b-d.

Across the $(\mathrm{Re}_{in},f)$-subspaces, the data stretches throughout the entire performance space for every case of $p$ considered (Figure \ref{fig:Results_COT_vs_Speed}b). On the other hand, the $(\mathrm{Re}_{in},p)$-subspaces divulge a slightly different story. As $f$ varies, the data still stretches throughout the entire performance space; however, it was more clustered. Higher $f$ resulted in faster speeds within the region of $10^3\lesssim COT \lesssim10^5$ (Figure \ref{fig:Results_COT_vs_Speed}c).
Furthermore, distinct localized clusters emerge in the $(f,p)$-subspace, for different values of $\mathrm{Re}_{in}$ (Figure \ref{fig:Results_COT_vs_Speed}d). Lower $\mathrm{Re}_{in}$ appear to be associated with generally slower speeds and higher $COT$ over every $(f,p)$-subspace considered. The cluster associated with $\mathrm{Re}_{in}=900$ appears near the region with minimal $COT$, centered about $1/\mathrm{St}=3$. These trends can be seen in specific subspaces in more detail in the Supplemental Materials; see Figures S5, S11, and S16. These figures present the data for a particular subspace using $1D$ parameterizations through the subspace itself, i.e., parameterizations of either $\mathrm{Re}_{in}, f$, or $p$ depending on the subspace investigated. Furthermore, the dimensional analog of Figure \ref{fig:Results_COT_vs_Speed}, is provided as Figure S19 in the Supplemental Materials, where swimming speed and cost of transport are given in units of bodylengths/second and N/kg, respectively.

%
%
%
%

\section{Discussion}
\label{discussion}

Two-dimensional immersed boundary simulations were used to explore parameter subspaces of an idealized, simple swimming model. A total of 6,357 simulations were performed using The College of New Jersey's high-performance computing cluster \cite{TCNJ:ELSA}, each requiring approximately 24 hours of computing time, resulting in $\sim152,568$ computational hours necessary to explore such parameter subspaces. Multiple swimming performance metrics were extensively quantified across broad subspaces given by combinations of two parameters $(\mathrm{Re}_{in},f)$, $(\mathrm{Re}_{in},p)$, and $(f,p)$ and $3$ specific values of third. Some of the general trends observed were:
\begin{enumerate}
    \item Higher swimming speeds $(1/St)$ occur for parameter subspaces involving higher $\mathrm{Re}_{in}$ ($\mathrm{Re}_{in}\gtrsim250$) and lower $f$ ($f\lesssim1.75\ \mathrm{Hz})$. These regions also corresponded to swimmers with high emergent peak-to-peak stroke amplitudes.
    \item These swimming speeds also correspond to Strouhal numbers (St) in the range in which the majority of swimming and flying animals are observed, i.e., $0.2<St<0.4$ \cite{Taylor:2003}.
    \item The subspaces in which produce greater swimming speeds (for higher $\mathrm{Re}_{in}$), also tend to require $COT$ roughly 2-10x higher than where it is minimal. 
    \item The highest non-dimensional swimming speeds (1/St) occurred over subspaces involving $\mathrm{Re}_{in}$ due to minimal peak-to-peak amplitudes. In units of bodylengths/stroke or bodylengths/second, these corresponding parameter combinations resulted in the slowest swimmers.
    \item Swimming performance was less affected by variations in the kinematic parameter, $p$, compared to variations in $f$ or $\mathrm{Re}_{in}$ for their respective ranges considered.
\end{enumerate}

While these trends may have been found without having performed such large numbers of simulations, even by possibly two orders of magnitude, our parameter explorations were able to divulge robust parameter subspaces that offer greater (or lesser) performance. For example, the simple swimmer propelled itself forward the slowest for $\mathrm{Re}_{in}\approx10$, even for a variety of $f$ and $p$. On the other hand, a robust subspace for high swimming speeds appeared to be for $\mathrm{Re}_{in}\gtrsim500$, $1\leq f\leq1.75\ \mathrm{Hz}$, and $p$ in any of the values sampled ($0.075<p<0.425$). Moreover, from the parameter subspace offering high swimming speeds, one could find further subspaces in which $COT$ was above (or below) particular thresholds. 

Overall, our study was also able to hint towards the performance metrics' underlying sensitivity to parameters, although a proper quantitatively-based global sensitivity analysis, such as Sobol Sensitivity Analysis \cite{Sobol:2001}, is still required. However, such sensitivity analyses require themselves a large number of simulations to be performed ($\sim10^3-10^4$ for $3$ parameters), in which parameter combinations $(\mathrm{Re}_{in},f,p)$ are selected through Sobol Sequences \cite{Saltelli:2002,Saltelli:2010}, rather than uniform sampling as carried out here. Thus, finding robust parameters subspaces in which reveal higher swimming performance may not be as trivial to find. It remains unclear whether after Sobol sampling from a $3D$ space and performing simulations, if the data could be projected into a subspace, i.e., project the $3D$ swimming performance metric data onto a $2D$ subspace, e.g., $(\mathrm{Re}_{in},f), (\mathrm{Re}_{in},p)$, or $(f,p)$, and offer as much insight into swimming behavior.

There exist other methods to explore parameter subspaces which attempt to reduce the dimensionality of the system and thus the overall necessity of having to explore such a large parameter space (and perform a seemingly infeasible number of computationally expensive simulations). One such method is called \textit{active subspaces} which finds large variations in the gradient of a model's output in order to construct a response surface in a lower dimensional space \cite{Russi:2010,Constantine:2015}. It has been successfully applied to numerous problems, including optimizing the design of an aircraft wing, where a low dimensional subspace was found within a 50-dimensional input parameter space. The low dimensional subspace that was found effectively described the variability within the lift and drag coefficients to such an extent that it revealed global trends within the original higher dimensional parameter space. This allowed for an efficient method to design an optimal wing \cite{Lukaczyk:2014}. Such an approach could prove beneficial while exploring the fitness landscapes of numerous mechanical systems, each composed of a high dimensional parameter space, along with their inherent global sensitivities to parameters, all in conjunction with possible convergent evolutionary processes. Although evolution does not optimize towards a global optima, it could be useful technique for identifying and analyzing trends across a variety of mechanical systems.

%
%
%
%

\bibliographystyle{naturemag}
\bibliography{Swim}

%
%
%
%

\begin{notes}[Acknowledgements]
The author would like to thank Lindsay Waldrop and Jonathan Rader for the invitation to participate in the SICB Symposium \textit{Melding Modeling and Morphology: integrating approaches to understand the evolution of form and function symposium} at the 2020 annual meeting. He would also like to thank Christina Battista, Karen Clark, Jana Gevertz, Laura Miller, Matthew Mizuhara, Emily Slesinger, and Lindsay Waldrop for comments and discussion. He also wishes to thank the anonymous reviewers for their careful reading of the manuscript and their very insightful, constructive feedback that significantly strengthened the manuscript. Computational resources were provided by the NSF OAC \#1826915 and the NSF OAC \#1828163. Support for N.A.B. was provided by the TCNJ Support of Scholarly Activity Grant, the TCNJ Department of Mathematics and Statistics, and the TCNJ School of Science.
\end{notes}

\end{document}